\documentclass[twocolumn,table]{aastex62}

\usepackage{amsmath,amssymb,amstext}
\usepackage{mathtools}
\usepackage{gensymb}
\usepackage{comment}
\usepackage{todonotes}

\usepackage{graphicx}
\usepackage{multirow}
\usepackage{chngcntr}
\usepackage{booktabs}
\usepackage{xcolor}
\usepackage{soul}

\newcommand{\iso}[2]{\ensuremath{^{#2}{\rm #1}}}
\newcommand{\nickel}{\ensuremath{^{56}{\rm Ni}}}
\newcommand{\mej}{\ensuremath{M_{\rm ej}}}
\newcommand{\vej}{\ensuremath{v_{\rm ej}}}
\newcommand{\ekin}{\ensuremath{E_{\rm k}}}
\newcommand{\mni}{\ensuremath{M_{56}}}
\newcommand{\mrp}{\ensuremath{M_{r \rm p}}}
\newcommand{\msun}{\ensuremath{M_{\odot}}}
\newcommand{\xni}{\ensuremath{\psi_{56}}}
\newcommand{\xrp}{\ensuremath{\psi_{r \rm p}}}

\newcommand{\grb}{GRB 211211A}
\newcommand{\rp}{\emph{r}-process}

\newcommand{\transient}{T211211A}
\newcommand{\lgrb}{lGRB}
\newcommand{\sgrb}{sGRB}

\defcitealias{Barnes.Metzger_2022.ApJL_rproc.collapsars}{BM22}
\defcitealias{Rastinejad.ea_2022Nature_grb.211211a.knova}{R22}

\shorttitle{GRB 211211A From A Collapsar?} 
\shortauthors{Barnes \& Metzger}

\begin{document}

\title{A collapsar origin for GRB 211211A is (just barely) possible}

\author[0000-0003-3340-4784]{Jennifer Barnes}
\affil{Kavli Institute for Theoretical Physics, Kohn Hall, University of California, Santa Barbara, CA 93106, USA}
\email{jlbarnes@kitp.ucsb.edu}
\author[0000-0002-4670-7509]{Brian D. Metzger}
\affil{Department of Physics and Columbia Astrophysics Laboratory, Columbia University, New York, NY 10027, USA}
\affil{Center for Computational Astrophysics, Flatiron Institute, 162 5th Ave, New York, NY 10010, USA} 

\begin{abstract}
Gamma-ray bursts (GRBs) have historically been divided into two classes. 
Short-duration GRBs are associated with binary neutron-star mergers (NSMs), while long-duration bursts are connected to a subset of core-collapse supernovae (SNe).
\grb{} recently made headlines as the first long-duration burst purportedly generated by an NSM.
The evidence for an NSM origin was excess optical and near-infrared emission consistent with the kilonova observed after the gravitational wave-detected NSM GW170817.
Kilonovae derive their unique electromagnetic signatures from the properties of the heavy elements synthesized by rapid neutron capture (the \rp) following the merger.
Recent simulations suggest that the ``collapsar’’ SNe that trigger long GRBs may also produce \rp{} elements.
While observations of \grb{} and its afterglow ruled out an SN typical of those that follow long GRBs, an unusual collapsar could explain both the duration of \grb{} and the \rp-powered excess in its afterglow.
We use semianalytic radiation transport modeling to evaluate low-mass collapsars as the progenitors of \grb{}-like events.
We compare a suite of collapsar models to the afterglow-subtracted emission that followed \grb{}, and find the best agreement for models with high kinetic energies and an unexpected pattern of \nickel{} enrichment.
We discuss how core-collapse explosions could produce such ejecta, and how distinct our predictions are from those generated by more straightforward kilonova models.
We also show that radio observations can distinguish between kilonovae and the more massive collapsar ejecta we consider here.
\end{abstract}
\keywords{Supernovae: core-collapse supernovae --- Nucleosynthesis: \emph{r}-process --- Gamma-ray bursts}

\section{Introduction} \label{sec:intro}

The durations of gamma-ray bursts (GRBs) follow a bimodal distribution, with short (\sgrb) and long (\lgrb) varieties \citep{Kouveliotou.ea_1993.ApJ_grb.two.classes}.
Observations have tied these two classes of ultrarelativistic jets
to distinct progenitors, with \lgrb s arising from a subset of highly kinetic core-collapse supernovae \citep[CCSNe; e.g.][]{Galama98_bwDisc} and \sgrb s originating in compact binary mergers \citep{Abbott.ea_2017ApJ_gw170817.grb}.

However, analyses of GRB populations \citep[e.g.][]{Zhang.Choi_2008.AandA_swift.grb.t90s,Tarnopolski_2015.AandA_fermi.grb.distribution} indicate overlap between the distributions of the durations that characterize each class, raising the spectre of GRBs whose timescales are outliers among bursts triggered by the same progenitor \citep[e.g.][]{Bromberg.ea_2013.ApJ_grb.t90.overlap}.  

While a few \lgrb s with no obvious associated SNe have been tentatively attributed to a non-SN progenitor \citep{DellaValle.ea_2006.Natur_grb.060614.no.sn,GalYam.ea._2002_02ap.data,Fynbo.ea_2006.Natur_grbs.without.sne}, the uncertain nature of the electromagnetic (EM) counterparts to compact binary mergers impeded the definitive association of these bursts with mergers.  
Nevertheless, it was suggested that these ``hybrid'' sGRB/lGRB events were related to a subclass of bursts whose light curves exhibited sGRB-like prompt spikes followed by temporally extended variable X-ray emission lasting tens or hundreds of seconds \citep[e.g.][] {Norris.Bonnell_2006.ApJ_grbs.extended.emiss,Perley.ea_2009.apj_grb.080503.extended.emiss}.

Multi-messenger observations of the binary neutron-star merger (NSM) GW170817 improved this situation dramatically by confirming \citep{Goldstein.ea_2017.ApJ_grb.gw170817} the theorized \citep{Paczynski_1986_GRB,Eichler_1989_sGRBs,Narayan.ea_1992.ApJ_sgrb.merger.progen} association between mergers and \sgrb s and providing a detailed look at the merger's ``kilonova'' counterpart
\citep{Arcavi.ea_2017Natur_gw170817.lco.emcp.disc,
Chornock.ea_2017ApJ_gw.170817.em.red.spec,
Coulter.ea_2017Sci_gw.170817.emcp.disc,
Drout.ea_2017Sci_gw.170817.emcp.disc,
Evans.ea_2017Sci_gw.170817.em.blue.spec,
Kasliwal.ea_2017Sci_gw.170817.em.interp,
Kilpatrick.ea_2017.Sci_gw.170817.spectrum.opt.nir,
McCully.ea_2017ApJ_gw.170817.blue.spec,
Nicholl.ea_2017ApJ_gw.170817.blue.spec,
Shappee.ea_2017.Science_kn170817,
Smartt.ea_2017Natur_gw170817.empc.disc,
SoaresSantos.ea_2017ApJ_gw.170817.empc.decam.disc,
Tanvir.ea_2017.ApJL_gw170817.emcp.disc,
Valenti.ea_2017ApJ_gw.170817.emcp.disc}.

This allowed \citet{Rastinejad.ea_2022Nature_grb.211211a.knova} (henceforth \citetalias{Rastinejad.ea_2022Nature_grb.211211a.knova})
to connect the recent \grb{} to an NSM \citep[see also][]{Troja.ea_2022.Nature_grb.211211a.nsm} despite its long duration: a $T_{90}$ of ${\sim}34$ s according to  
 the \emph{Fermi} Gamma-ray Burst Monitor \citep{Mangan.Fermi.Team_2021.GCN_grb211211a.fermi.gbm}, or ${\sim}51$ s, as measured by \emph{Swift}'s Burst Alert Telescope \citep{Stamatikos.ea_2021.GCN_grb211211a.swift.bat}.
The association was based on the similarity of the optical and near-infrared (NIR) transient that emerged after the burst to the kilonova that arose following GW170817, as well as on the GRB's extended emission, whose duration and spectral evolution mimicked those observed to follow some sGRBs \citep[e.g.,][]{Gompertz.ea_2022.NatAst_grb.211211a.nsm}.

In a variation on that theme, \citet{Yang.ea_2022.Nature_nswd.grb211211a} proposed that the progenitor of \grb{} was the merger of a white dwarf with a NS or stellar-mass black hole (BH), which produces an accretion disk as disrupted white dwarf material circularizes around the central remnant \citep{Fryer.Woosley_1998.ApJL_He.star.Bh.Grb.mod}.
However, this interpretation is in tension with semianalytic
\citep{Metzger_2012.mnras_wdns.disks,Margalit.Metzger_2016.mnras_wdns.merger.disk.nucleo,Kaltenborn.ea_2022.arXiv_nswd.nucleosyn} and numerical \citep{Fernandez.ea_2019.mnras_nuc.acc.flows.2,Zenati.ea_2019.mnras_nswd.mergers} simulations of these disks, which cast doubt on their ability to effectively neutronize, a precondition for \emph{r}-production.

The LIGO-Virgo gravitational-wave (GW) detector network was offline at the time of \grb, so no GW data were available to confirm a compact object merger coincident with the burst.
However, the position of the burst, offset 7.91 kpc from the center of its putative host galaxy \citepalias{Rastinejad.ea_2022Nature_grb.211211a.knova}, supports the merger theory, as compact object binaries receive kicks during the SN explosions of their component stars, and often travel far from their hosts' centers before they merge \citep[e.g][]{Kalogera.ea_1998.ApJ_sn.kicks.ns.bin}.
Some authors \citep[primarily][who propose an alternate, dust-based explanation for the NIR emission]{Waxman.ea_2022.arXiv_grb.211211a.nir.dust} have cast doubt on the host identification.
However, since a distance is required to determine the luminosity of the transient and make comparison to our models, we are unable to engage with the undiscovered-host hypothesis in this work.

Kilonovae are distinguishable by their uniquely red spectra, a hallmark imparted by the high opacities of select elements burned by rapid neutron capture (the \rp), a nucleosynthesis channel that operates in the neutron-rich gas formed from NS material unbound during the merger.

However, kilonovae may not be the only explosions in which the \rp{} occurs. 
General relativistic magnetohydrodynamic (GRMHD) simulations of the accretion disks that form in the CCSN explosions of rapidly rotating massive stars (``collapsars'') suggest that conditions in these disks can become neutron-rich \citep{Siegel.Barnes.Metzger_2019.Nature_rp.collapsar}, allowing the \rp{} to synthesize heavy elements in winds blown off the disk. 
While not all simulations of collapsar disks predict a robust \rp{} in disk outflows \citep{Miller.ea_2020.ApJ_rproc.collapsar.blue,Just.ea_2022.mnras_nuetrino.cooled.bh.acc.disks,Fujibayashi.ea_2022.arXiv_collapsars.disk.winds}, the \rp{} collapsar hypothesis is also supported by patterns in Galactic chemical evolution that seem to require an \rp{} source that tracks star formation \citep{Cote.ea_2019.ApJ_other.rproc.sources,Naidu.ea_2022.ApJ_disrupted.halos.rproc.sources}.
(A short delay time characterizes CCSNe in general, but is harder to square with NSMs, which represent the endpoint of an evolutionary track that unfolds over hundreds of millions or even billions of years \citep[e.g.][]{Belczynski.ea_2002.ApJ_pop.synth.compact.binaries}.)

Collapsars were originally proposed to explain \lgrb s and the high-velocity, broad-lined Type Ic (Ic-BL) SNe that often accompany them.
The implication then is that \emph{r}-production may coincide with GRBs regardless of their duration.

We investigate here the possibility that \grb{} was triggered by a collapsar, and that its optical and NIR counterpart, which we label as a transient of undetermined classification, \transient{}, is the emission from an \rp-enriched SN, albeit a unique one.
We describe our semi-analytic radiation transport scheme, and the models to which we apply it, in \S\ref{sec:methods}.
In \S\ref{sec:results}, we present the models that best reproduce the emission of \transient{}, and discuss their properties.
We explore in \S\ref{sec:discussion} what subclass of collapsars might be able to produce these properties, but ultimately fail to convince ourselves that such explosions represent a superior explanation for 
\transient. 
We also outline how radio observations can distinguish between the low-mass collapsar progenitors we focus on and the more conventional kilonova explanation for \transient.
We leave our parting thoughts in \S\ref{sec:conclusion}.

\section{Methods}\label{sec:methods}

We use a semianalytic radiation transport model to predict the emission from \rp-enriched collapsars with a variety of parameters, which we compare to observations of \transient.

\subsection{Radiation Transport Model}\label{subsec:rtmodel}
We repurpose the radiation transport framework developed by \citet{Barnes.Metzger_2022.ApJL_rproc.collapsars}
(hereafter \citetalias{Barnes.Metzger_2022.ApJL_rproc.collapsars}), in which the SN ejecta is divided into concentric shells whose internal energies evolve in response to radioactive heating, adiabatic expansion, and the diffusion and free-streaming of radiation.
A full discussion of the implementation can be found in \citetalias{Barnes.Metzger_2022.ApJL_rproc.collapsars}.
Here, we highlight minor adjustments we have made to our previous models and methods, which better position us to study the apparently low-mass and high-velocity explosion \citepalias{Rastinejad.ea_2022Nature_grb.211211a.knova} that produced \transient.
 
First, we no longer assume that \nickel{} is evenly distributed in the ejecta.
The ejecta configurations we consider are described in more detail in \S\ref{subsec:modeldesc}.
For consistency, when calculating the $\gamma$-ray opacity to determine the deposition of \iso{Ni/Co}{56} decay energy \citep[\`{a} la][]{Colgate.ea_1980.ApJ_sn.grays.deposition.lum}, we now include only the ejecta layers that contain \nickel.

We also now explicitly account for the thermalization of \rp{} decay products beyond $\gamma$-rays.
Our current models have lower masses and higher velocities than the \rp-enriched SNe of \citetalias{Barnes.Metzger_2022.ApJL_rproc.collapsars}.
The resulting lower densities reduce the optical depth for thermalizing interactions \citep{Barnes_etal_2016}, rendering suspect the assumption of efficient thermalization of $\beta^-$- and $\alpha$-particles and fission fragments. 
We adopt the approximate analytic formula for thermalization efficiency $f_{\rm th}^{\rm rp}$ from \citet{Barnes_etal_2016},
\begin{equation*}
f_{\rm th}^{r \rm p} = 0.36\left(\exp[-0.55 t_{\rm d}] + \frac{\ln[1+0.26t_{\rm d}^{0.9}]}{0.26t_{\rm d}^{0.9}}\right),
\end{equation*}
where $t_{\rm d}$ is the time in days, and we have chosen coefficients corresponding to kilonovae with \rp{} masses and velocities most similar to those of our low-mass collapsar models.
This factor is applied to a baseline \rp{} heating rate $\dot{Q}_{r \rm p} = 2.0 \times 10^{10} \; t_{\rm d}^{-1.3}$ erg s$^{-1}$  g$^{-1}$ \citep[e.g.][]{Metzger_2010,Korobkin_NSM_rp}. 

Finally, the short rise time of \transient{} motivates an explicit accounting of the thermal energy deposited in the ejecta during the explosion. 
(In typical GRB-SNe, which rise to peak ${\sim}$1--2 weeks after explosion, and which burn larger quantities of \nickel{} \citep{Prentice.ea_2016_SeSne.bolLCs,Taddia.ea_2019.AandA_IcBL.iPTF.survey,Perley.ea_2020.ApJ_ZTF.sn.sample}, energy from \nickel{} decay rapidly dominates the adiabatically degrading initial thermal energy, preventing the thermal component from influencing the light curve.)

We assume there is a characteristic time, $t_{\rm eq}$, at which the thermal and kinetic energy in a given ejecta layer are in equipartition. 
The subsequent conversion of the former to the latter accelerates each layer to its final kinetic energy.
By the time the SN light curve becomes visible, this conversion is effectively complete; though thermal energy remains, it is insufficient to alter the ejecta's velocity structure. 
Thus, it is valid to approximate the initial thermal energy as equal to half the final kinetic energy in ejecta shell $i$, $E_{{\rm k},i}$.
The residual energy at $t_0$, the start time of the simulation, is then
\begin{align}
    E_{{\rm th},i} = \frac{1}{2}E_{{\rm k},i} \left(\frac{t_0}{t_{\rm eq}}\right)^{-1}.\label{eq:eth_init}
\end{align}

The models of \citetalias{Rastinejad.ea_2022Nature_grb.211211a.knova} also include a thermal component, which they attribute to a cocoon created by the GRB jet as it burrows through the ejecta.
In the collapsar scenario, $E_{{\rm th},i}$ could be the product of an initial supernova explosion.
It could  also result from a shock interaction that occurs when the eventual accretion disk wind collides with either the SN ejecta or (in the case of an temporally accelerating disk outflow; see Sec.~\ref{subsec:viability}) with itself.

In the interest of limiting the dimensionality of our model suite,  we do not treat $t_{\rm eq}$ as a free parameter.
However, preliminary explorations found that $t_{\rm eq} = 1$ s allowed us to fit the early blue and ultraviolet (UV) emission. 
This value should be treated as a rough indicator---the exact balance that is achieved between thermal and kinetic energy and whether that balance is uniform over the entire ejecta, for example, are open questions.
Nevertheless, it points to heating timescales that could be compatible with either jet breakout or a prompt explosion.

\subsection{Model Suite}\label{subsec:modeldesc}

Our model suite is summarized in Table~\ref{tab:modelsuite}.
Based on the arguments of \citetalias{Rastinejad.ea_2022Nature_grb.211211a.knova}, we focus on collapsar models with low masses and high velocities. 
We consider total ejecta masses in the range $0.5\msun \leq \mej \leq 1.0\msun$, and average ejecta velocities $\vej$ of 0.1--0.35$c$, where $\vej = \sqrt{2\ekin/\mej}$, with \ekin{} the ejecta's kinetic energy.
In all our models, mass density follows a broken power law, $\rho(v) \propto v^{-d}$, with $d=1$ (10) in the inner (outer) parts of the ejecta.
The low luminosities of \transient,
relative to the GRB-SN population, suggest lower quantities of \nickel, so we restrict our exploration to models with $0.01\msun \leq \mni \leq 0.1\msun$.

We consider \rp{} masses $\mrp/\msun$ of 0.0, 0.02,  0.05, and 0.08.
These values were motivated by the luminosity of \transient, which constrains the total radioactive mass to be low.
That they are lower than what was suggested by \citet{Siegel.Barnes.Metzger_2019.Nature_rp.collapsar} for typical collapsar \rp{} yields (${\lesssim}1 \msun$) also reflects the overall lower ejecta masses in this work. 
(In contrast, \citet{Siegel.Barnes.Metzger_2019.Nature_rp.collapsar} focused on the more massive progenitors proposed by \citet{Heger.Woosley_2000.ApJ_massive.rotating.stars} to explain CCSNe with higher \mej.)

As mentioned in \S\ref{subsec:rtmodel}, our ejecta structure is more complex than in \citetalias{Barnes.Metzger_2022.ApJL_rproc.collapsars}, since the \nickel{} mass fraction is no longer required to be uniform.
Instead, we extend \nickel{} from some inner normalized mass coordinate \xni{} to the edge of the ejecta.
Such a configuration might be realized if \rp{} winds fail to mix completely with the earlier ejecta containing whatever \nickel{} is burned by the prompt explosion.
As in \citetalias{Barnes.Metzger_2022.ApJL_rproc.collapsars}, the \rp{} material is mixed from the center of the ejecta out to a normalized mass coordinate \xrp.

Given \mej, \mni, and \mrp{}, the quantities \xni{} (\xrp{}) can take on values from $0$ to $[1-\mni/\mej]$ ($\mrp/\mej$ to $1$).
For each parameter combination, we choose five values of \xni{} and \xrp{} that are spaced uniformly within the ranges defined above.
We assume that \nickel{} (\rp{} material) is evenly distributed over $m_{\rm enc} \geq \xni$ ($m_{\rm enc} \leq \xrp$), and consider all combinations of \xni{} and \xrp{} for which the sum of \nickel{} and \rp{} mass fractions is less than or equal to unity everywhere in the ejecta. 
As in \citetalias{Barnes.Metzger_2022.ApJL_rproc.collapsars}, the opacity of an ejecta shell is determined by its composition. 
Ejecta lacking both \nickel{} and \rp{} elements is assumed to have a baseline opacity of 0.05 cm$^2$ g$^{-1}$.

\begin{table}[h]
\begingroup
\centering
\caption{Parameters of the model suite}\label{tab:modelsuite}
    \begin{tabular}{p{0.1\columnwidth}>{\raggedright\arraybackslash}p{0.36\columnwidth}>{\raggedleft\arraybackslash}p{0.42\columnwidth}}
    \toprule
     \emph{Symbol} & \emph{Definition}  & \emph{Values} \tabularnewline
    \hline
    \mej{} & Total ejecta mass & 0.5\msun{} -- 1.0\msun, $\Delta\mej/\mej = 0.08$ \\
    \vej & Average ejecta velocity & $0.1c$ -- $0.35c$, $\Delta\vej/\vej = 0.18$ \\
    \mni & \nickel{} mass & 0.01\msun{} -- 0.1\msun, $\Delta \mni = 0.01$ \\
    \mrp & \emph{R}-process mass & (0.0, 0.02, 0.05, 0.08)\msun  \\
    \xni{} & Lowest mass coordinate with \nickel{} & 0 -- ($1-\mni/\mej$), $\Delta \xni = (1-\mni/\mej)/5$ \\
    \xrp & Highest mass coordinate with \rp{} matter & $(\mrp/\mej)$ -- 1, $\Delta \xrp = (1-\mrp/\mej)/5$ \\
    \bottomrule
    \end{tabular}
    \endgroup
\end{table}

\subsection{Model Evaluation}\label{subsec:model_eval}

We calculate the broadband evolution of our model in $ugriz$, $B$, $J$, and $K$ bands for every combination of the parameters delineated in Table~\ref{tab:modelsuite}, and compare the results to the afterglow-subtracted photometry of \transient{} published in \citetalias{Rastinejad.ea_2022Nature_grb.211211a.knova}, for times ${\geq}0.05$ days.

We quantify the agreement between the data and each instantiation of the model using a simple chi-square metric,
\begin{align*}
    \chi^2 &= \sum_{i} \frac{(F_{{\rm obs},i} - F_{{\rm pred},i})^2}{\sigma_{i}^2} \\
    &+ \sum_{j} \frac{\left[\max(F_{{\rm pred},j} - F_{{\rm ul},j}, 0)\right]^2}{\sigma_{\rm est}},
\end{align*}
where $F_{{\rm obs}, i}$ ($F_{{\rm pred},i}$) is the observed (predicted) flux corresponding to measurement $i$, which we derive from reported magnitudes, and $\sigma_{\rm i}$ is the uncertainty on the $i$th measurement.
The second sum runs over reported upper limits, $\{F_{{\rm ul},j}\}$.
Its terms contribute to $\chi^2$ only when the model's predicted flux exceeds the upper limit. 
The variable $\sigma_{\rm est}$ is an estimated uncertainty on the upper limit, which we set to 0.1 mag.

\section{Results}\label{sec:results}

\begin{figure}\includegraphics[width=\columnwidth]{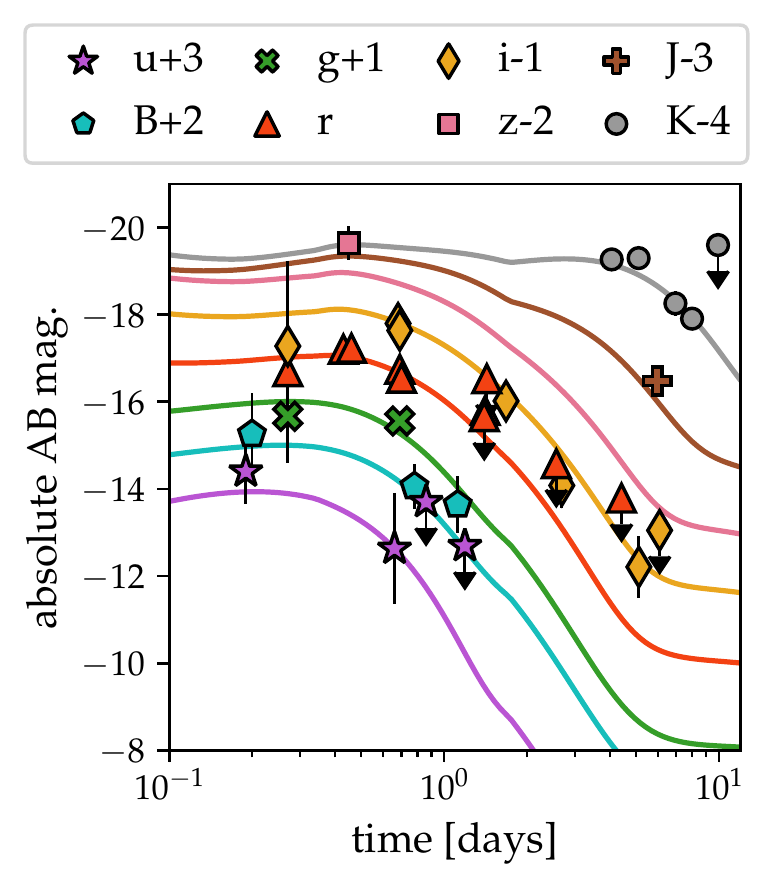}
\caption{The model from our suite with the lowest $\chi^2$ has $\mej = 1.0 \msun$, $\vej = 0.26c$, $\mni  = 0.01\msun$, $\mrp = 0.05\msun$, $\xni = 0.99$ and $\xrp = 0.76$.
While these parameters provide a good fit to the data, they also define a rather extreme ejecta configuration, in which radioactive \nickel{} is concentrated in a shell at the outer edge of the ejecta.}\label{fig:best-fit}
\end{figure}

We perform a grid search to locate the model in the suite with the lowest $\chi^2$, and find that the best match to the data (with $\chi^2 \approx 32$) is achieved by the parameters $\mej = 1.0 \msun$, $\vej = 0.26c$, $\mni  = 0.01\msun$, $\mrp = 0.05\msun$, $\xni = 0.99$, and $\xrp = 0.76$.
The light curve for this model is compared to data in Fig.~\ref{fig:best-fit}.

While this model agrees well with the data, degeneracies among the parameters and the simplicity of the semianalytic model motivate us to investigate additional ejecta models.
Furthermore, our procedure does not circumscribe the distribution of \nickel{} in the ejecta beyond the physical requirement that $(1-\xni) \mej \geq \mni$. 
The model above, which features an outer shell composed of pure \nickel, is allowed within our framework.
However, it is worth determining whether less extreme ejecta configurations can reproduce the data with comparable fidelity.
In \S\ref{subsec:successprops}, we zoom out and identify larger populations of models with a range of parameters that nonetheless provide good matches to the photometry of \transient.

\subsection{Properties of successful models}\label{subsec:successprops}

Before presenting predictions generated by particular parameter combinations, we briefly survey the landscape of all models that provide a satisfactory fit to the observations. 
We define a satisfactory fit as one for which $\chi^2 \leq 100$.
Since our model has six degrees of freedom ($N_{\rm dof} = 6$) and is fit against 40 observations and upper limits, this translates to a reduced chi-square metric $\chi^2_{\rm red} \equiv \chi^2/N_{\rm dof} \lesssim 2.5$.
This filter selects ${\sim}1600$ models, or just over 2\% of the full suite.

\begin{figure}\includegraphics[width=\columnwidth]{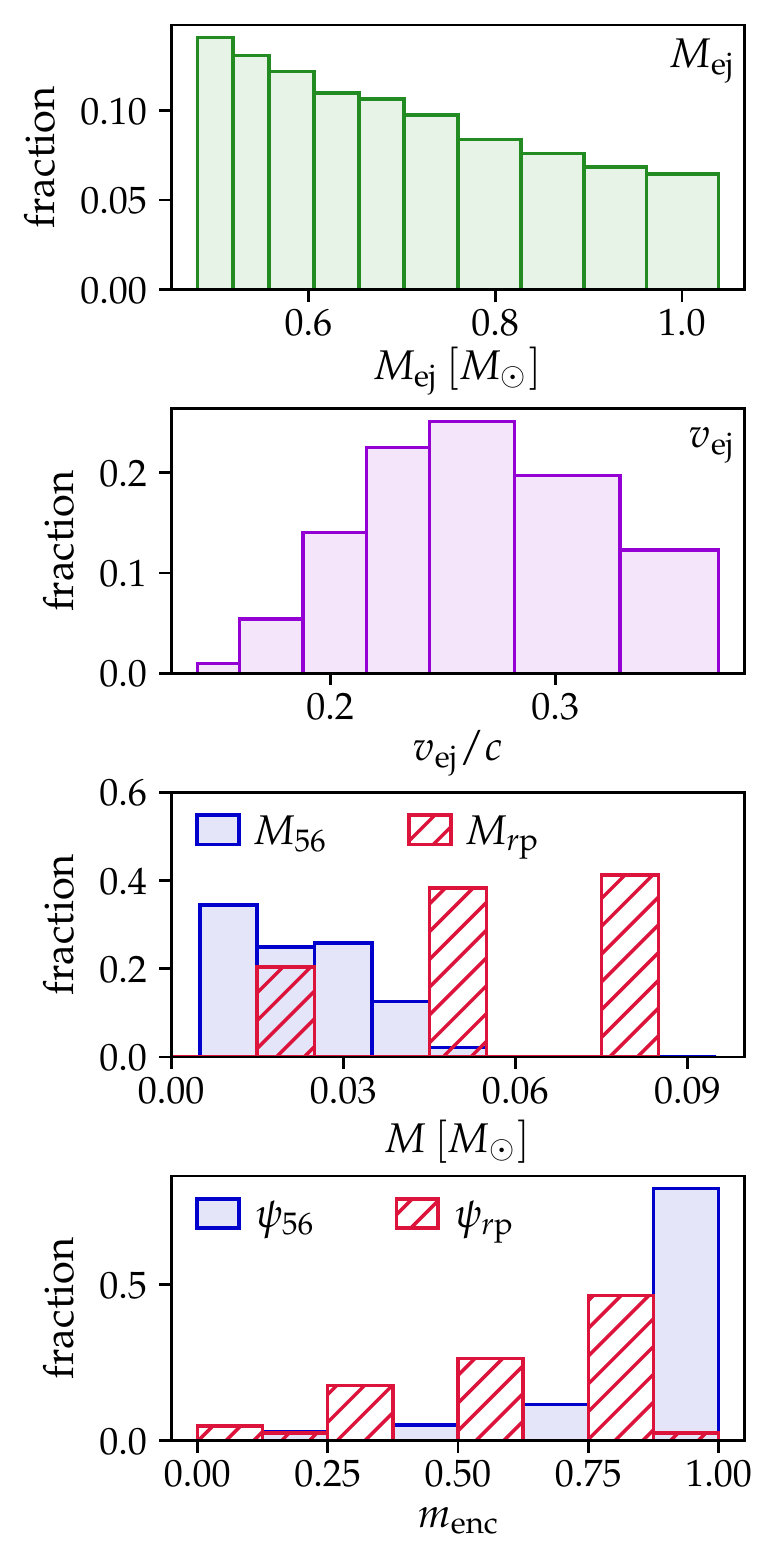}
\caption{The distribution of model parameters for models with $\chi^2 \leq 100$.
The good-fitting models span the full range of \mej{} in our model suite (\emph{first panel}), but draw primarily from the upper end of our \vej{} range ($\vej \gtrsim 0.2c$; \emph{second panel}).
While various \rp{} masses, $0.01\msun \leq \mrp \leq 0.08\msun$, can be compatible with the observations, lower \nickel{} masses ($\mni \lesssim 0.05 \msun$) are preferred (\emph{third panel}).
The majority of the successful models (\emph{fourth panel}) feature well-mixed \rp{} material, but concentrate their \nickel{} in a thin shell at the outer edge of the ejecta.
In the top two panels, the variable widths of the bars reflect the logarithmic spacing of the model parameters.
}\label{fig:model_hist}
\end{figure}

Fig.~\ref{fig:model_hist} shows how the six model parameters are distributed within the good-fitting model set.
Models with good fit scores draw from the full range of \mej{} we consider, though they evince a slight preference for lower ejecta masses.
The range of velocities is narrower; agreement with the data is easier to achieve for $\vej \gtrsim 0.2c$.
While such velocities are similar to those inferred for the kilonova model of \citetalias{Rastinejad.ea_2022Nature_grb.211211a.knova}, when combined with low-mass collapsars' larger ejecta masses (vis-\`{a}-vis kilonovae), they imply kinetic energies near or beyond the upper limit of what has historically been considered possible for SNe \citep{Thompson.ea_2004.ApJ_magnetar.sne.grb,Mazzali2014_Magnetars,ChenEtAl2017_BlicMagnetar}.

The parameters governing \rp{} and \nickel{} production and distribution complete the picture.
As the third panel shows, all \rp{}  masses we consider (except $\mrp = 0$, which cannot produce the observed NIR excess) can yield photometry more or less consistent with observations.
Masses of \nickel{} are more tightly constrained; none of the good-fitting models have $\mni > 0.05 \msun$.

As indicated in the final panel, the majority of the good-fitting models feature a particular mixing pattern, in which \rp{} material is mixed out from the center to fairly high normalized mass coordinates $m_{\rm enc}$, while \nickel{} is concentrated in the outermost layers of the ejecta.
We will discuss in \S\ref{sec:discussion} if this configuration is strictly necessary to reproduce the photometry of \transient, and whether an outflow with such a radially stratified composition could be produced in nature.

\subsection{Successful model clusters}\label{subsec:clusters}
To better understand how successful models are situated within
the six-dimensional parameter space in which our suite is defined, we use the Agglomerative Clustering routine of Python's \texttt{scikit-learn} package \citep{scikit-learn} to sort them into five groups.
The hierarchical clustering algorithm in the \texttt{SciPy} library guided our choice of the number of clusters.

The coordinates of the cluster centroids are reported in Table~\ref{tab:centroid_params}, along with the percentage of good-fitting models belonging to each cluster.
These data provide additional insight into the combinations of parameters capable of reproducing the photometry of \transient.

\begin{table}[h]
\begingroup
\centering
\caption{Cluster centroids of the successful models}\label{tab:centroid_params}
    \begin{tabular}{>{\centering\arraybackslash}p{0.2\columnwidth}>{\centering\arraybackslash}p{0.07\columnwidth} p{0.07\columnwidth}>{\centering\arraybackslash}p{0.08\columnwidth}>{\centering\arraybackslash}p{0.08\columnwidth}>{\centering\arraybackslash}p{0.07\columnwidth}>{\centering\arraybackslash}p{0.07\columnwidth}}
    \toprule
     \emph{Index} (\%)  & $\mej^{\dag}$ & $\vej^{\ddag}$ & $\mni^{\dag}$ & $\mrp^{\dag}$ & \xni{} & \xrp{} \tabularnewline
    \hline
    1 \hspace{2 mm} (19)  & 0.61 & 0.25 & 0.036 & 0.053 &  0.94 & 0.68 \\    
    2 \hspace{2 mm} (27)  & 0.65 & 0.23 & 0.022 & 0.067 & 0.97 & 0.33 \\
    3 \hspace{2 mm} (19)  & 0.60 & 0.31 & 0.012 & 0.053 & 0.60 & 0.75 \\
    4 \hspace{2 mm} (19)  & 0.88 & 0.22 & 0.024 & 0.053 & 0.97 & 0.66 \\
    5 \hspace{2 mm} (15)  & 0.63 & 0.30 & 0.016 & 0.049 & 0.97 & 0.64 \\
    \bottomrule
    \end{tabular}
    \endgroup
    \vspace{0.5mm}
    \newline \footnotesize{$^{\dag}$ Values in units of \msun.}
    \newline \footnotesize{$^{\ddag}$ Values in units of $c$.} 
\end{table}

While some of the cluster centroids share the combination of high \ekin{} and extreme \xni{} suggested by Fig.~\ref{fig:model_hist}, Table~\ref{tab:centroid_params} shows that these characteristics are not required to reproduce the data within our error tolerance.
In fact, aside from centroid 5, all the centroids differ from the best-fit model in at least one significant way.
Of particular interest are centroid 1, which has barely half the kinetic energy of the best-fit model; centroid 2, which has both lower \ekin{} and a lower \xrp; and centroid 3, which has more extensive \nickel{} mixing.
Still however, Table~\ref{tab:centroid_params} suggests some trade-off between \xni{} and \vej. 
Successful models with more extensive \nickel{} mixing have higher average velocities.
This is required to reproduce the light curves' rapid evolution; a spatially extended emitting region must expand faster to yield a similar light-curve time scale. 

\begin{figure*}\includegraphics[width=\textwidth]{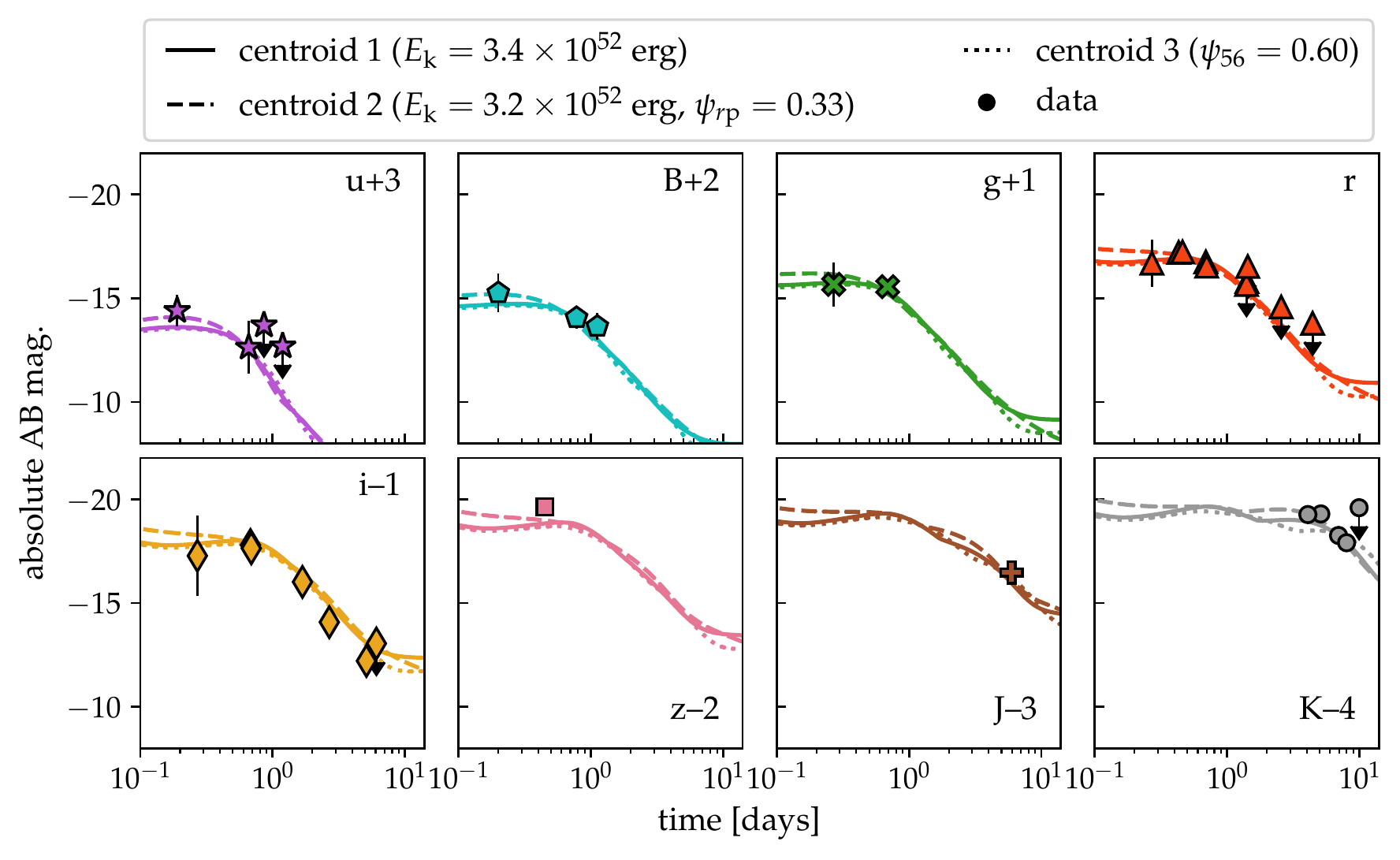}
\caption{Due to degeneracies among model inputs, diverse sets of parameters produce comparable light curves.
The panels above show the broadband light curves for some of the centroids defined in Table~\ref{tab:centroid_params}, which differ from the best fit model either in their level of \nickel{} or \rp{} mixing, or in their kinetic energy.
Data from \citetalias{Rastinejad.ea_2022Nature_grb.211211a.knova} are shown for comparison.
While very large \ekin{} and minimal \nickel{} mixing are common to many of the good-fitting models, they are apparently not required to reproduce the data.
}\label{fig:select-centroids}
\end{figure*}

In Fig.~\ref{fig:select-centroids}, we show the light curves produced by the centroids (1, 2, and 3) highlighted above.
While the agreement with observations is by definition poorer than for the best-fit model, each set of parameters reproduces the fundamental characteristics of \transient.
Given the simplicity of our radiation transport method, the only slightly poorer fits are not sufficient reason to discard these models.

\section{Discussion}\label{sec:discussion}

As explained in \S\ref{subsec:clusters}, due to degeneracies among parameters, low-mass collapsar models with varying physical properties reproduce the photometry of \transient{} with comparable fidelity.
However, even these degeneracies do not allow infinite flexibility; all of the models have very high velocities and/or poorly mixed \nickel{} that would render them outliers among observed GRB-SNe and SNe Ic-BL.
We next discuss two possible interpretations of these results, and outline how radio observations can distinguish low-mass collapsars from standard kilonovae.

\subsection{A low-mass collapsar?}\label{subsec:viability}

The low ejecta masses we explore here, which are necessitated by the swift evolution of \transient, are already a departure from the standard collapsar picture, in which a few solar masses of stellar material are ejected \citep[e.g.][]{Cano.ea_2017_SN.GRB.Review}.
The formation of an accretion disk---the defining feature of the collapsar model---is enabled by the rapid rotation of the pre-explosion star.
Processes that remove mass from the star earlier in its evolution (e.g., line-driven winds or stripping by a companion) also siphon away the angular momentum that allows disk formation.
Our low-\mej{} models thus correspond more naturally to a scenario in which a large fraction of the pre-explosion mass is captured by the NS or BH formed during the explosion than one in which the progenitor mass is unusually low at the point of collapse.

The low masses and modest \nickel{} production that characterize our good-fitting models could plausibly arise from the explosion of a star with slightly less angular momentum than in more typical collapsars \citep[e.g.][]{Janiuk.Proga_2008.ApJ_lowL.accretion.grb.time,Murguia-Berthier.ea_2020.ApJL_stellar.rot.bh.transient}.
The proto-neutron star produced when such a progenitor collapses \citep[e.g.,][]{Dessart.ea_2008.ApJL_pns.collaspar.long.soft.grb} would initially rotate relatively slowly.  This, coupled with the delay between the initial collapse and the circularization of the outer layers into an accretion disk, may preclude the kind of prompt (${\lesssim}1$ second post-collapse) MHD jetted explosion \citep[e.g.,][]{Mosta.ea_2014.ApJL_magnetorot.sne.3d,Varma.ea_2021.mnras_mhd.sn.sim} invoked to explain the copious \nickel{} production in more typical SNe Ic-BL \citep[e.g.][though see \citet{Zenati.ea_2020.mnras_nuclear.burning.acc.disks} for an alternative \nickel{} production site]{Barnes.Duffell.ea_2018.ApJ_grb.icbl.engine}.
A weaker explosion could nonetheless launch a low-mass outflow enriched with \nickel{} burned in the inner layers~\citep{Maeda_nucleosynth_jetdriv},
thus forming the outer layers of the SN ejecta. 

Subsequent material would be ejected once the infalling material had coalesced into an accretion disk.
While most of the disk mass would accrete onto the central remnant, powering a relativistic jet \citep[e.g.][]{Bromberg.Tchekhovskoy_2016.mnras_rel.mhd.ccsn.grb}, a fraction would become gravitationally unbound and expand outward at mildly relativistic velocities \citep[e.g.][]{Siegel.Metzger_2017.PRL_nsm.accretion.disks.rproc}.

The accretion rate onto the disk will decline with time, with consequences for nucleosynthesis in the winds.
Early high accretion rates through the disk support cooling by neutrino emission \citep{De.Siegel_2021.ApJ_NSM.accretion.disks}, followed by the neutronization of the disk mid-plane \citep{Siegel.Barnes.Metzger_2019.Nature_rp.collapsar,Just.ea_2022.mnras_nuetrino.cooled.bh.acc.disks,Fujibayashi.ea_2022.arXiv_collapsars.disk.winds}.
If the newly neutron-rich matter from the mid-plane escapes the disk without re-protonizing, an \rp{} can occur as it decompresses upon ejection.
As the accretion rate drops, neutronization of the infalling material ceases, truncating \emph{r}-production in disk outflows.
Thereafter, disk winds are composed of He and, to a much lesser degree, iron-peak elements formed in the disk-wind outflows when the electron fraction  $Y_{\rm e} \approx 0.5$ \citep{Siegel.Barnes.Metzger_2019.Nature_rp.collapsar,Zenati.ea_2020.mnras_nuclear.burning.acc.disks}.
These later ejections account for the non-radioactive mass in our ejecta models. 

The time-dependent disk-outflow properties also depend on the strength and structure of the magnetic field feeding the BH.  
The early, \rp-rich winds are likely ejected with velocities ${\sim}0.1c$, corresponding to a weak poloidal magnetic field \citep{Siegel.Metzger_2017.PRL_nsm.accretion.disks.rproc}.
However, subsequent outflows may be launched at increasingly high velocities, as continual accretion strengthens the magnetic field in the disk \citep[e.g.][]{Tchekhovskoy.Giannios_2015.MNRAS_mag.flux.star.grb,Gottlieb.ea_2022.arxiv_bh.to.photosphere}.
For a sufficiently strong and ordered poloidal magnetic flux, wind velocities could reach ${\approx}0.3c$ \citep{Christie.ea_2019.MNRAS_bfield.geom.nsm.discs}.

The higher velocity of the later-stage ejecta would induce mixing between disk-wind outflows launched at different times and substantively increase the ejecta's total kinetic energy.  
Such velocity evolution can therefore account both for the high average velocities and the compositional profiles of the good-fitting models.
However, the general lack of evidence for \nickel-mixing means that the earliest mass ejection must occur at velocities high enough to avoid mixing with the disk-wind matter.

We see that with a modest degree of fine-tuning, this scenario can explain the fundamental features of our favored ejecta models.
We emphasize that this ejecta configuration is likely to differ from a garden-variety (higher angular momentum) collapsar, for which the overall ejecta mass is larger and a greater fraction of the disk outflows may be \rp-enriched, due to the higher accretion rates at early times.  

\subsection{A collapsar in kilonova clothing?}\label{subsec:interp}

While we argued in \S\ref{subsec:viability} that nature may produce ejecta similar to those described in \S\ref{subsec:successprops}, a more skeptical reading of our analysis is that it selects models whose emission is fundamentally similar to a kilonova.

The two traits that distinguish our low-mass collapsars from kilonovae are the production, albeit limited, of \nickel{} and the significant quantities of non-\rp{} ejecta.
However, our good-fitting models have ejecta configurations that dampen the effects of these attributes on their emission, relative to comparable kilonova models. 

The low mass of \nickel, combined with its position at high velocities, limits its impact on the resulting SN.
Of the relatively little energy produced by \nickel{} decay, only a small fraction is thermalized, due to the low densities near the outer edge of the ejecta where the \nickel{} is located \citep{Colgate.ea_1980.ApJ_sn.grays.deposition.lum}.
What energy does thermalize diffuses rapidly through the low-optical-depth layers at the ejecta's edge.
Its effects are ephemeral, and easily overpowered by the signal from the ejecta's residual thermal energy
(Eq.~\ref{eq:eth_init}).

\begin{figure}
\includegraphics[width=\columnwidth]{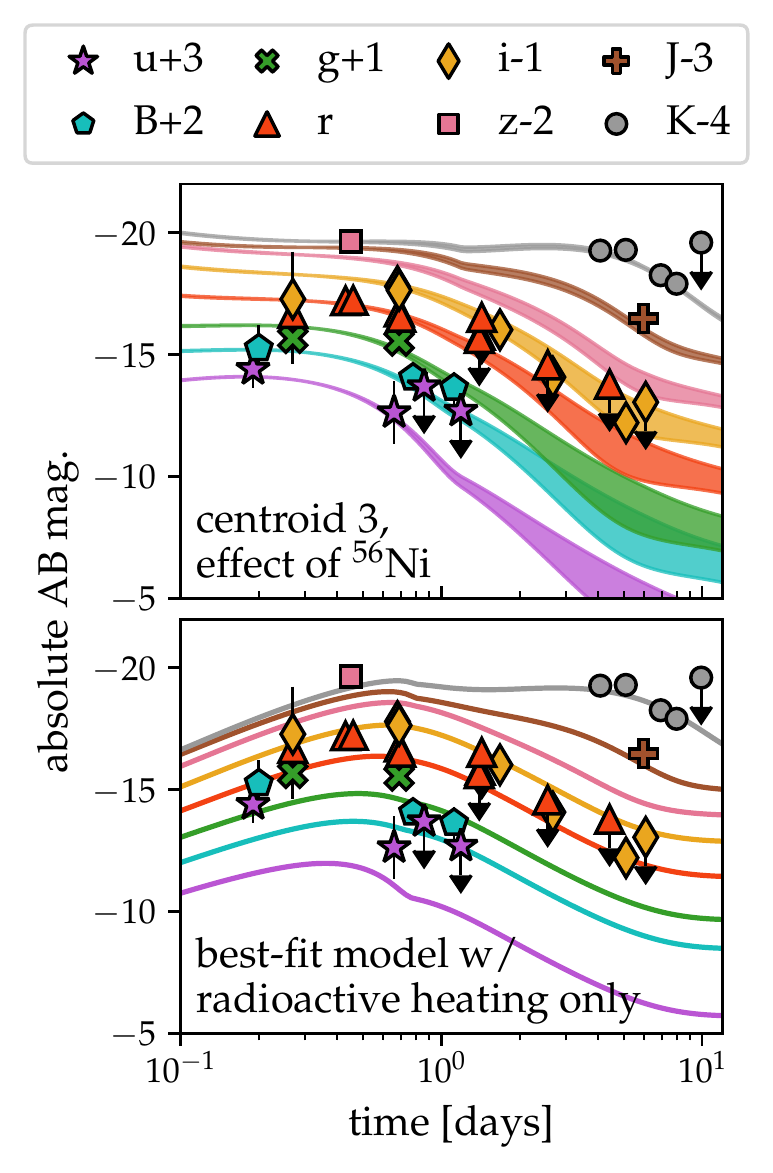}
\caption{Modified low-mass collapsar models probe the effects of \nickel{} and on the emission. 
In both panels, we compare to data from \citetalias{Rastinejad.ea_2022Nature_grb.211211a.knova}.
\emph{Top panel}: Broadband light curves for an unaltered centroid 3 model and a version with $\mni=0$, which form the upper and lower bounds of the filled curves, respectively.
Removing \nickel{} does not fundamentally change the emission; the apparent differences at $t \gtrsim 1$ day are due mainly to our assumptions about emission from optically thin ejecta.
\emph{Bottom panel:}
The best-fit model (with $\mni = 0.06 \msun$ and $\xni = 0.9$) from a suite in which heating is due solely to radioactivity fails to match the early signal, suggesting that \nickel-heating is not a substitute for $E_{{\rm th},i}$.
The minor role of \nickel{} in our original good-fitting models is not due to our inclusion of an initial thermal energy reservoir (Eq.~\ref{eq:eth_init}), but instead reflects the incompatibility of the early data with copious, well-mixed \nickel.}
\label{fig:ni_v_jet}
\end{figure}

The relative invisibility of \nickel{} in our models is illustrated in the top panel of Fig.~\ref{fig:ni_v_jet}, which shows the impact of removing \nickel{} on the light curves of the centroid 3 model (Table~\ref{tab:centroid_params} and Fig.~\ref{fig:select-centroids}).
We selected centroid 3 because its \nickel{} is mixed more thoroughly into the ejecta than in other centroids, which should increase the sensitivity of the emission to \nickel{} decay.  Although the model including \nickel{}, whose light curves form the upper bounds of the shaded curves in Fig.~\ref{fig:ni_v_jet}'s top panel, is brighter than the model without, whose light curves constitute the lower bounds, these differences become most significant at later times, when the data are less constraining, and modeling efforts face more uncertainties (e.g., the nature of optically thin emission; see \citetalias{Barnes.Metzger_2022.ApJL_rproc.collapsars}).
The effect at $t \lesssim 1$ day is minimal, because at these times the radiation of residual thermal energy dominates. 

To test whether our assumption of an initial thermal component biases our analysis against models with larger \mni{} or lower \xni, we run a separate model grid with the same parameter ranges defined in Table~\ref{tab:modelsuite}, but which omits $E_{{\rm th},i}$ as defined by Eq.~\ref{eq:eth_init}.
Instead, we initialize the internal energies of the ejecta shells by estimating the combined effects of radioactive heating and adiabatic expansion for $t \leq t_0$, which results in much lower internal energies.

The bottom panel of Fig.~\ref{fig:ni_v_jet} shows the light curves of the best-fit model from this grid, which has $\mej = 0.58 \msun$, $\vej = 0.3c$, $\mni = 0.06\msun$, $\mrp = 0.08\msun$, $\xni = 0.90$, and $\xrp = 0.78$.
Its $\chi^2$ is 86, higher than the best-fit model in our original suite, but comparable to the models in our good-fitting subset. 
While \mni{} is slightly higher than in the original good-fitting model subset (see Fig.~\ref{fig:model_hist}), the \nickel{} is again concentrated in the ejecta's exterior.
This suggests that the \nickel{} in our original suite was not forced to the edge of the ejecta by our adopted model for $E_{{\rm th},i}$, but rather that significant and/or well-mixed \nickel{} decreases agreement with observations.
In particular, \nickel{}, on its own, cannot explain the earliest emission, particularly in bluer bands.
Given that \nickel{} is not necessary to explain the late-time signal (see top panel) and appears to be insufficient to explain the earlier parts of the light curves, we conclude that \nickel{} is \emph{allowed} but not \emph{required} by the data.

The position of \nickel{} in our good-fitting models also calls into question the import of the ejecta's non-radioactive material.
With \nickel{} restricted to the outermost layers, the outward diffusion of the energy from \nickel-decay is effectively independent of \mej.
While energy from \rp{} decay must diffuse through a larger fraction of the ejecta, the opacity it encounters is dominated by \rp{} elements; the low opacity of the inert material means its effect on diffusion times is minimal.
Thus, though non-radioactive matter dominates \mej, its influence on the emission may be subtle.

\begin{figure}\includegraphics[width=\columnwidth]{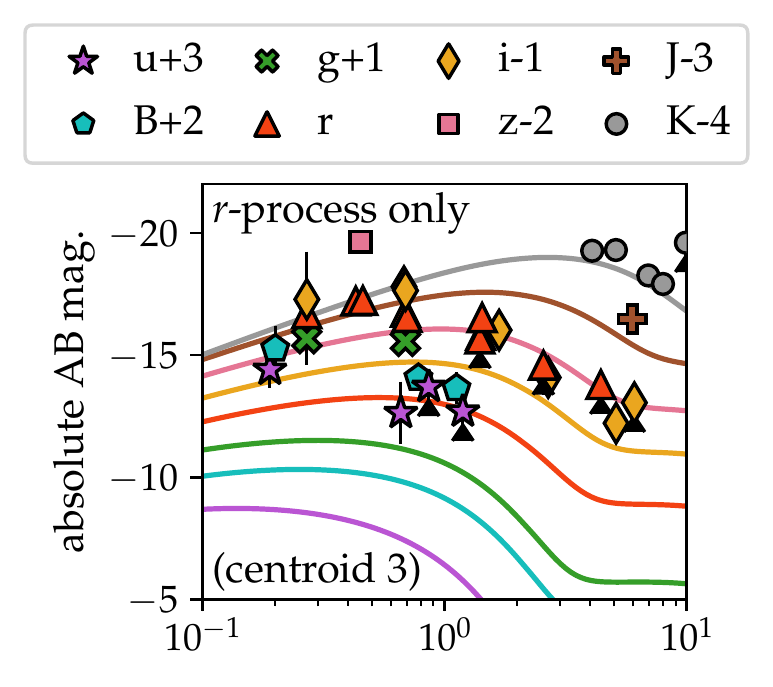}
\caption{Broadband light curves for a kilonova model containing only the \rp{} ejecta from centroid 3, compared to photometry from \citetalias{Rastinejad.ea_2022Nature_grb.211211a.knova}.
Non-\rp{} material is not required to explain the NIR emission of \transient.
However (see text), the lower masses of kilonovae compared to collapsars require different assumptions about the initial thermal energy in order to match the earliest and bluest observations.
}\label{fig:c6_rproc}
\end{figure}

To explore the role of non-radioactive material, we transform the centroid 3 collapsar model into a kilonova by excising all of its non-\rp{} ejecta. 
(I.e, this model has $\mej = \mrp = 0.053\msun$ and a reduced $\vej = 0.26c$ on account of its lower mass.)
Our adopted \rp{} opacity ($\kappa_{r \rm p} = 10$ cm${^2}$ g$^{-1}$) means this pure \rp{} model corresponds to a kilonova that originated in low-$Y_{\rm e}$ conditions and is rich in lanthanides and actinides. 
In other words, its composition is akin to that of a ``red'' kilonova \citep[e.g][]{Barnes_2013}.
The model's light curves are displayed in Fig.~\ref{fig:c6_rproc}.

The agreement in $J$ and $K$ remains decent, confirming that non-radioactive matter has only a small impact on radiation from the \rp-enriched layers.
The largest effect is on the early signal, particularly in bluer bands, which suffers because of a reduction in the initial internal energy resulting from the reduced mass of the kilonova ejecta.
($E_{{\rm th},i}$ scales with shell mass in our model; see Eq.~\ref{eq:eth_init}.)

While we do not attempt to optimize a kilonova model here, our current method for determining $E_{{\rm th},i}$ suggests such disagreement would be robust across a broad range of kilonova parameters, owing to the vastly different mass scales of kilonovae and even low-mass collapsars.
However, kilonova models with added complexity can avoid the early-time disagreement.
\citetalias{Rastinejad.ea_2022Nature_grb.211211a.knova} achieved a good fit to the observations by incorporating two additional, lower-opacity (hence bluer) kilonova components, as well as a shock-heated cocoon.

In an echo of our earlier discussion of \nickel{}, we conclude that large quantities of non-radioactive mass are neither ruled out by the data nor necessary to explain them. 

\subsection{Tie-breaker: radio emission}\label{subsec:radio}

Our analysis does not conclusively favor a low-mass collapsar origin for \transient.
However, the possibility remains that it, or a future transient with similar properties, could be generated by a collapsar explosion with the combination of parameters detailed in \S\ref{subsec:modeldesc}.
In the event that nature conspires to produce such an explosion, its late-time radio signal could offer a way to distinguish it from a kilonova born of an NSM.

Both collapsars and kilonovae generate synchrotron radio emission as their ejecta collide with material surrounding the explosion site and decelerate.
The rise of the resulting radio light curve, which takes anywhere from a few to several years, is related to the distribution of the fastest material, and therefore sensitive to assumptions about the density profile at the edge of the ejecta. 
In contrast, the eventual light-curve peak reflects the total kinetic energy contained in the ejecta, which is greater for energetic low-mass collapsars than for mergers by more than an order of magnitude, due principally to the higher masses of the former.
Because of the recentness of \grb{}, radio non-detections obtained since the burst, like those of \citetalias{Rastinejad.ea_2022Nature_grb.211211a.knova}, most strongly constrain the high-velocity tail of the ejected matter.
Continued observations will be invaluable for probing the total kinetic energy of the explosion.

Following \citet{Nakar.Piran_2011.Nature_radio.bns.mergers} and \citet{Kathirgamaraju.ea_2019.mnras_knova.radio.afterglows}, we estimate the properties of the radio signals from collapsars and kilonovae. 
The time at which the radio light curve peaks is
\begin{align}
    t_{\rm pk} \approx 2.7 \text{ yr}  \left(\frac{E_{51}}{\beta_0^2}\right)^{1/3} \left(\frac{3}{5\beta_0}-1\right),\label{eq:tpk_radio}
\end{align}
where $E_{51}$ is the kinetic energy of the explosion in foe, $\beta_0$ is the velocity of the slowest ejecta layer, and we have eliminated the dependence on the circumburst number density by fixing $n$ to the value reported in \citetalias{Rastinejad.ea_2022Nature_grb.211211a.knova} ($n=0.54$).

If we additionally adopt the values \citetalias{Rastinejad.ea_2022Nature_grb.211211a.knova} derived for the fractions of energy in electrons ($\epsilon_{\rm e} = 3.28 \times 10^{-2}$) and magnetic fields ($\epsilon_{\rm B} = 1.52 \times 10^{-4}$), we can estimate the peak flux at a given radio frequency $\nu$ as
\begin{align}
    F_{\nu, \rm pk } \approx 26 \: \mu\text{Jy } E_{51} \beta_0^{\frac{5p-7}{2}} \nu_{9.5}^{\frac{1-p}{2}}.\label{eq:peak_flux}
\end{align}
In Eq.~\ref{eq:peak_flux}, $\nu_{9.5}$ is $\nu$ normalized to $10^{9.5}$ Hz and \citetalias{Rastinejad.ea_2022Nature_grb.211211a.knova}'s value of $p=2.014$ is used to calculate the prefactor (and for consistency should be adopted when evaluating the exponents).
We have also converted from luminosity to flux assuming the distance to \transient{} is 350 Mpc \citepalias{Rastinejad.ea_2022Nature_grb.211211a.knova}.

\begin{figure}\includegraphics[width=\columnwidth]{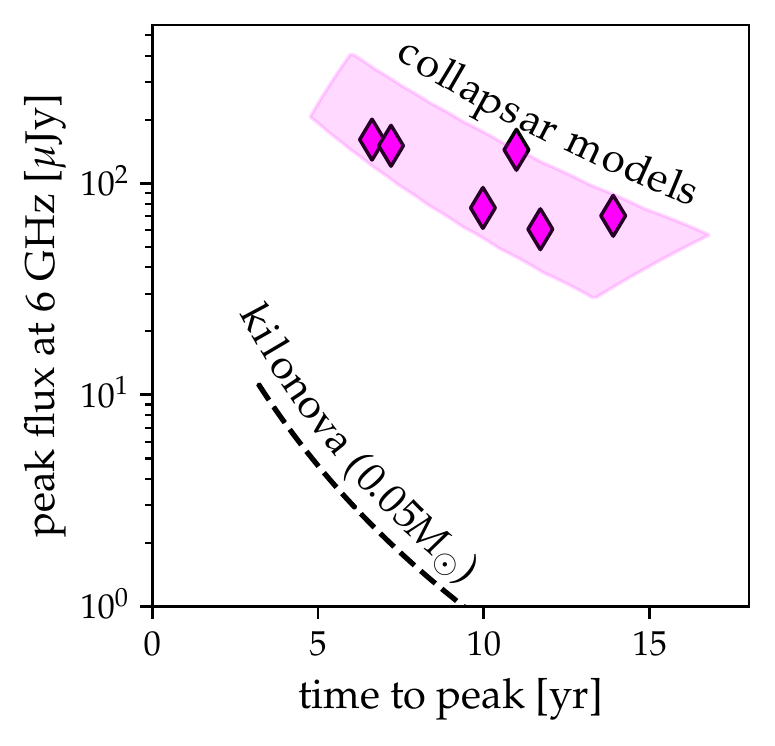}
\caption{Even low-mass collapsars generate much brighter radio afterglow emission than kilonovae. 
The time-to-peak and peak flux at 6 GHz for our best-fit collapsar model and the centroid parameters of Table~\ref{tab:centroid_params} are plotted as pink diamonds.
The pink shaded region shows the expected peak properties for collapsars with masses (velocities) in the range $0.5\msun$--$1.0\msun$ ($0.2c$--$0.35c$),
which typify the properties of our good-fitting models. 
We show as a dashed black line the peak properties of kilonovae with $m_{\rm ej,k} = 0.05\msun$ and $0.1c \leq v_{\rm ej,k} \leq 0.3c$.
}\label{fig:radio_peaks}
\end{figure}

Fig.~\ref{fig:radio_peaks} shows the peak time and peak flux at 6 GHz of our best-fit collapsar model and our five centroids, calculated according to Eqs.~\ref{eq:tpk_radio} and \ref{eq:peak_flux} and assuming that $\beta_0 = \vej/c$ for each model.
(The exact value of $\beta_0$ is difficult to define for realistic ejecta density profiles, but since the fastest moving layers of the ejecta carry the majority of the kinetic energy, this choice is reasonable.)
The pink shaded region in Fig.~\ref{fig:radio_peaks} shows the range of peak properties for collapsars with parameters that this work suggested might produce emission consistent with \transient: $0.5\msun \leq \mej \leq 1.0\msun$ and $0.2c \leq \vej \leq 0.35c$.

For comparison, we also plot the peak properties of a kilonova with $m_{\rm ej,k} = 0.05\msun$ (the total \rp{} mass suggested by \citetalias{Rastinejad.ea_2022Nature_grb.211211a.knova}) and a range of velocities $0.1c \leq v_{\rm ej,k} \leq 0.3$ as a dashed black line. 
(We consider a range of velocities because the minimum velocity is nontrivial to define in the case of a multi-component model like the one constructed in \citetalias{Rastinejad.ea_2022Nature_grb.211211a.knova}.)
Due to their greater kinetic energies, the collapsar models have much higher fluxes at peak than a kilonova would when the parameters beyond  $E_{51}$ and $\beta_0$ are held constant.
Ground-based radio telescopes (e.g., the Very Large Array) could easily distinguish between these cases with long-term monitoring of the radio signal.

\section{Conclusion}\label{sec:conclusion}

We have used semianalytic radiation transport modeling to investigate the possibility that the ambiguous \grb{} originated not in a compact object merger, as proposed by \citetalias{Rastinejad.ea_2022Nature_grb.211211a.knova}, \citet{Troja.ea_2022.Nature_grb.211211a.nsm} and \cite{Yang.ea_2022.Nature_nswd.grb211211a}, but rather in the CCSN explosion of a star with less angular momentum than a typical lGRB progenitor.
According to this theory, the \rp{} elements that provide the NIR excess observed in the GRB afterglow were synthesized not from neutron-rich material expelled during the coalescence of a neutron-star binary, but from ordinary stellar material that became neutron-rich in an accretion disk mid-plane as a result of weak interactions in the presence of electron degeneracy \citep{Siegel.Barnes.Metzger_2019.Nature_rp.collapsar}.
Our model assumes that the expulsion of this material from the disk enriches the central core of the SN ejecta with \rp{} elements.

We find that certain regions of our parameter space produce emission that broadly agrees with observations of the afterglow-subtracted light curves of \transient.
However, the particular constellations of parameters required to achieve a reasonable fit---namely very high velocities and the presence of \nickel{} only at the outer edges of the ejecta---point to an explosion distinct from the standard picture of collapsars.

Further bedeviling the interpretation of \transient{} is the fact that \nickel---at least when restricted to the ejecta's edge---has only a minor impact on the emission. 
The large quantity of non-radioactive material (the other feature that distinguishes our low-mass collapsars from the merger-driven models of \citetalias{Rastinejad.ea_2022Nature_grb.211211a.knova} and \citet{Yang.ea_2022.Nature_nswd.grb211211a}) plays a larger role, but its importance is contingent on our assumptions about how internal energy is generated in the earliest phases of the explosion.
Equally plausible treatments put forward by \citetalias{Rastinejad.ea_2022Nature_grb.211211a.knova} are able to account for the early blue emission without appealing to mass beyond the \rp{} material required to explain the NIR excess.
Thus we conclude that although a collapsar could explain \transient, nothing about \transient's emission serves as a smoking gun for a collapsar progenitor.

Fortunately, the lack of clarity surrounding \grb{} and its afterglow will itself be transient. 
We have shown that radio observations can easily distinguish signals produced by collapsar ejecta from those generated by the much less massive outflows produced by merging compact objects.
Furthermore, in the future, gravitational-wave detectors will definitively settle the question of a merger v. collapsar trigger for difficult-to-classify GRBs.
In the multi-messenger era, we can hope to understand the full diversity of GRB emission and progenitors.

\section{Acknowledgments}

The authors thank A. Polin, J. Rastinejad, G. Schroeder, and A. V. Villar for helpful conversations.
J.B. gratefully acknowledges support from the Gordon and Betty Moore Foundation through Grant GBMF5076
B.D.M. is supported in part by the National Science Foundation (Grants AST-2009255, AST-2002577). 
This work was performed in part at Aspen Center for Physics, which is supported by National Science Foundation grant PHY-1607611,
as well as the  Kavli Institute for Theoretical Physics at the University of California at Santa Barbara, which  receives funding from the National Science Foundation though Grant PHY-1748958.

\bibliographystyle{apj} 
\bibliography{refs}

\end{document}